\documentclass{article}

\PassOptionsToPackage{numbers}{natbib}

 \usepackage[preprint]{neurips_2022}

\usepackage[utf8]{inputenc} 
\usepackage[T1]{fontenc}    
\usepackage{hyperref}       
\usepackage{url}            
\usepackage{booktabs}       
\usepackage{amsfonts}       
\usepackage{nicefrac}       
\usepackage{microtype}      

\usepackage{amsmath}
\usepackage{algorithm2e}

\usepackage{multirow}
 \usepackage{graphicx}
 \usepackage[table,xcdraw]{xcolor}
 \usepackage{multirow}
\usepackage{multicol}
\usepackage{caption}
\usepackage{subcaption}
\usepackage{comment}
\title{On the Impact of Quantization and Pruning of Self-Supervised Speech Models for Downstream Speech Recognition Tasks “In-the-Wild”}

\author{%
\hspace{-1.2cm}
Arthur Pimentel$^1$ \quad Heitor Guimarães$^1$ \quad Anderson R. Avila$^2$ \quad Mehdi Rezagholizadeh$^2$ \quad Tiago H. Falk$^1$ \\
$^1$Institut National de la Recherche Scientifique \quad $^2$Huawei Noah’s Ark Lab\\
\texttt{\{arthur.pimentel,heitor.guimaraes,tiago.falk\}@inrs.ca}\\
\texttt{\{anderson.avila,mehdi.rezagholizadeh\}@huawei.com}
}

\begin{document}

\maketitle

\begin{abstract}
Recent advances with self-supervised learning have allowed speech recognition systems to achieve state-of-the-art (SOTA) word error rates (WER) while requiring only a fraction of the labeled training data needed by its predecessors.  Notwithstanding, while such models achieve SOTA performance in matched train/test conditions, their performance degrades substantially when tested in unseen conditions. To overcome this problem, strategies such as data augmentation and/or domain shift training have been explored. Available models, however, are still too large to be considered for edge speech applications on resource-constrained devices, thus model compression tools are needed. In this paper, we explore the effects that train/test mismatch conditions have on speech recognition accuracy based on compressed self-supervised speech models. In particular, we report on the effects that parameter quantization and model pruning have on speech recognition accuracy based on the so-called robust wav2vec 2.0 model under noisy, reverberant, and noise-plus-reverberation conditions. 
\end{abstract}

\section{Introduction}
Large deep learning models have recently achieved great success on speech recognition tasks \cite{speechT5, speech_XLS-R}. These models, however, use a considerable amount of computational resources, which can be unfeasible for many edge applications. 
Edge applications focus on bringing computing as close to the source of data as possible in order to reduce latency and bandwidth use. It can be particularly important for speech recognition applications, where private and/or sensitive speaker data may need to be sent over the network to be processed remotely on large data processing clusters hosting very large and complex models. Bringing such large models to the edge can be challenging, as some edge devices may be resource-constrained with limited storage and processing capacity. Moreover, edge applications are corrupted by several environmental factors, such as ambient noise and/or room reverberation, which are known to be detrimental to speech-based applications. As such, a more detailed study on the impact of model compression and inference efficiency for large speech recognition models is needed. We aim to fill this gap.

More specifically, in this study our overarching goal is two-fold: (1) understand how well state-of-the-art (SOTA) speech recognition models behave under different model compression schemes, and (2) how well the compressed model accuracy remains under varying noise and reverberation conditions. We hope that the results from this study will shed light on the performance gaps that may exist before ``edge speech recognition'' is implemented in practice. Experiments with the latest (robust) wav2vec 2.0 model are conducted under two different compression schemes (quantization and model pruning) and five noisy conditions (SNR = 0, 5, 10, 15, and 20 dB) and two reverberations conditions (small room and medium room).

\section{Methods and Materials}
\subsection{Speech Recognition Models}
The \textit{wav2vec 2.0}~\cite{wav2vec2.0} model learns basic speech units used to tackle a self-supervised task. The architecture consists of a multi-layer convolutional feature encoder which takes as input raw audio and outputs latent speech representations at each time step, which are then fed to a context network. The encoder consists of several blocks containing a temporal convolution followed by layer normalization~\cite{layernorm2016} and a GELU~\cite{gelu2016} activation function, while the context network is a Transformer creating contextualised representations from the entire sequence.

The \textit{robust wav2vec 2.0}~\cite{robusst_wav2vec2.0} model is a recent variant developed to provide improved robustness against domain shifts (e.g., due to noise, varying datasets, or other factors) at test time. Using the same architecture as its predecessor, robust wav2vec utilizes target domain data during pre-training, thus leading to significant performance improvements in out-of-domain ASR. In its original proposal, \citet{robusst_wav2vec2.0} evaluated model performance in different out-of-domain conditions. However, varying noise levels -- an important condition in edge applications -- was not explored comprehensively. Here, we aim to fill this gap, as well as gauge the impact that model compression may have. In our experiments, pre-trained and fine-tuned models from the \textit{Hugging Face} platform were used. More specifically, the wav2vec 2.0 model is pre-trained and fine-tuned on 960 hours of the Librispeech dataset\footnote{\url{https://huggingface.co/facebook/wav2vec2-large-960h}}, while the robust wav2vec 2.0 is pre-trained on the Libri-Light, CommonVoice, Switchboard and Fisher datasets and fine-tuned on 960 hours of the Librispeech dataset\footnote{\url{https://huggingface.co/facebook/wav2vec2-large-robust-ft-libri-960h}}.


\subsection{Model compression techniques}
Two classic model compression techniques are explored to gauge the potential of edge speech recognition applications. The first is quantization, where the number of bits required to store each weight is reduced, thus substantially shrinking the model size, saving memory and accelerating computation. The method can be further extended to represent gradient and activation in the quantized form \cite{model_compression_survey_ieee}. Here, we explore the impact of 8-bit quantization on all linear layers of the speech models and compare against its original 32-bit version (i.e., a compression ratio of 4). Next, model pruning is explored where redundant parameters can be removed from the network with minimal effect on model accuracy~\cite{Choudhary2020}. Global unstructured pruning based on the lowest L1-norm was used at five different pruning rates, from 10-30\% at 5\% intervals.

\subsection{Additive noise and reverberation}
As we are interested in understanding the impact of compressed speech models in edge conditions, we use the noise signals present in the Deep Noise Suppression Challenge 4 (DNS4) dataset ~\cite{dubey2022icassp} to corrupt the test speech signals of the Librispeech dataset. This noise dataset consists of 180 hours of noise, present across 62,000 utterances, covering  150 different non-speech-like noise types. These files are added to the test samples at five varying SNR levels, ranging from 0 dB to 20 dB at 5 dB intervals.
Reverberation, in turn, is simulated by convolving the clean signals with a uniformly sampled room impulse response (RIR). The openSLR28 dataset with 248 real RIRs and the openSLR26 with 60,000 synthetic RIRs are used~\cite{ko2017study} to simulate small and medium sized rooms. Lastly, reverberation and noise is simulated by combining the two previous steps. In all cases, if necessary, waveforms are resampled to 16~kHz. 

\section{Results and Discussion}

\begin{figure}
    \centering
        \begin{subfigure}[htpb]{0.49\textwidth}
            \centering
            \includegraphics[width=\textwidth]{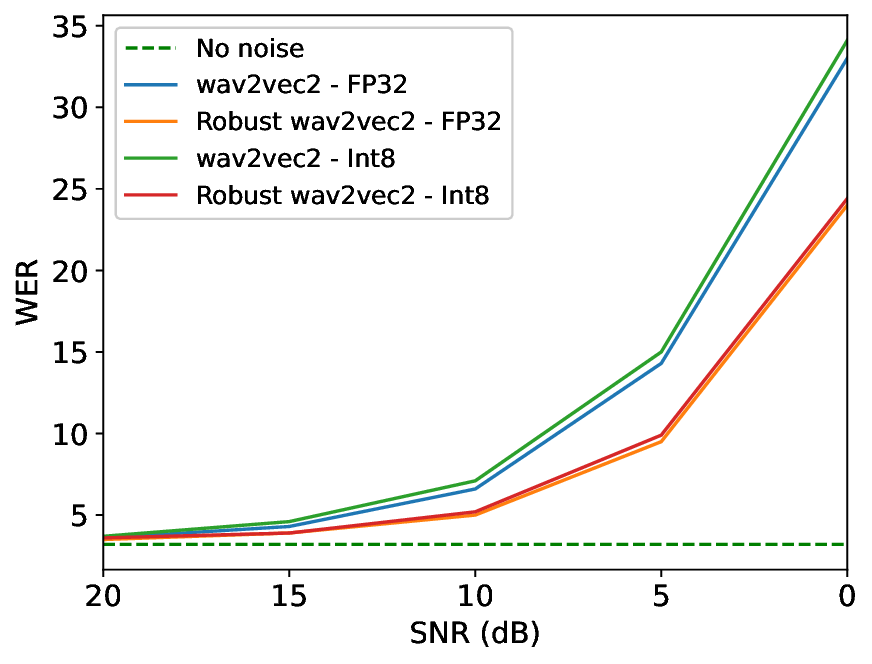}
            \caption{}
            \label{fig:quant_noise_robust}
        \end{subfigure}
        \begin{subfigure}[htpb]{0.49\textwidth}
            \centering
            \includegraphics[width=\textwidth]{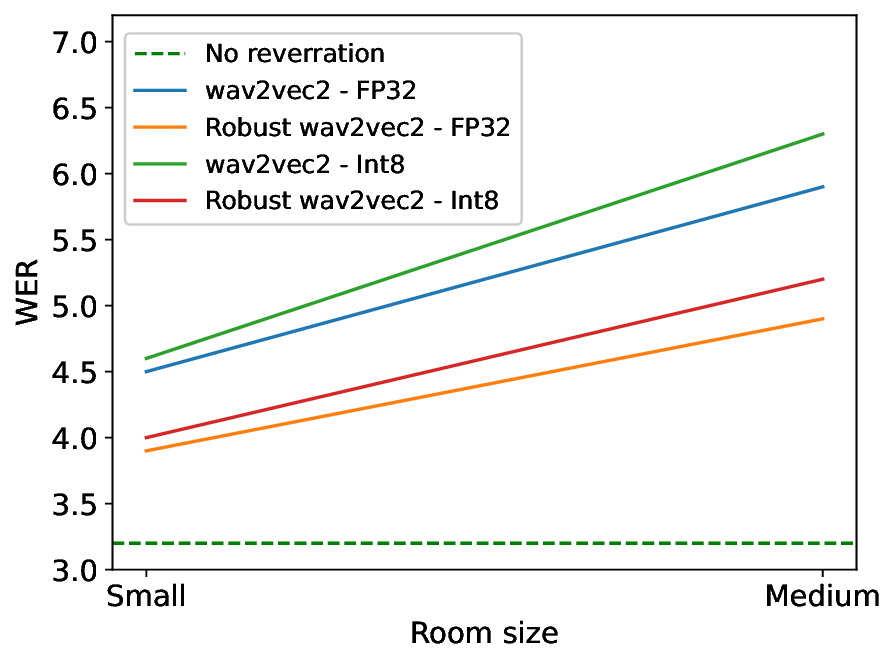}
            \caption{}
            \label{fig:quant_reverb_robust}
        \end{subfigure}
    \caption{WER for original (FP32) and quantized (Int8) models, with (a) noise and (b) reverberation.}
    \label{fig:prune_robust_noise_reverb}
    \end{figure}

First, we explore the impact of quantization on both original and robust versions of wav2vec 2.0 in clean matched conditions. Both models achieved a WER of 3.2\% with weights stored in 32-bit floating point precision (total model size of 1262 Mb). After quantization, WER increased to 3.3\%, thus a slight increase for a 4-fold model compression (total model size 354.5 Mb). Next, we explore the impact of pruning the weights of the convolutional and linear layers of the models. It is observed that the robust version of the speech model showed only a slight increase in WER to 3.3\% at a prune rate of 0.3, while its original version increased to 5.7\%. 

Next, we explore the impact of varying noise levels and reverberation. Figures \ref{fig:quant_noise_robust} and \ref{fig:quant_reverb_robust} show WER plots of original and quantized models, as a function of SNR and room size, respectively. As can be seen, 8-bit quantization showed minimal effect on model performance for the noisy case and a small impact when reverberation was present (e.g., WER for robust wav2vec went from 0.049 to 0.052 for the medium room size condition). Next, we perform an in-depth analysis of the WER achieved for different noise types. Table \ref{tab:wer_noise_type} shows the mean WER for audio files corrupted with six common noise types, namely domestic sounds (e.g., vacuuming), human voice, music, vehicle noise, wind noise, and other miscellaneous sources. As can be seen, human voice, domestic sounds, and vehicle noise showed the greatest performance deterioration. These are conditions in which edge applications would typically be seen, such as smart speakers and in-vehicle speech recognition. Overall, the robust wav2vec 2.0 model outperformed the original wav2vec 2.0 across all noise types, with the closest match achieved with wind noise.

\begin{table}[]
\centering
\caption{Mean WER per noise type.}
\label{tab:wer_noise_type}
\begin{tabular}{ccc}
\hline
\multirow{2}{*}{\textbf{Noise Type}} & \multicolumn{2}{c}{\textbf{Mean WER}}              \\ \cline{2-3}
                                 \\[-0.9em]
                                     & \textbf{wav2vec 2.0} & \textbf{Robust wav2vec 2.0} \\ \hline
                                     \\[-0.9em]
Domestic sounds                      & 13.6                & 10.6                       \\
Human voice                          & 12.0                & 9.3                       \\
Miscellaneous sources                & 6.3                & 3.3                       \\
Music                                & 11.0                & 7.6                       \\
Vehicle                              & 11.7                & 8.8                       \\
Wind                                 & 5.8                & 5.4                       \\ \hline
\end{tabular}
\end{table}

\begin{figure}
    \centering
        \begin{subfigure}[htpb]{0.49\textwidth}
            \centering
            \includegraphics[width=\textwidth]{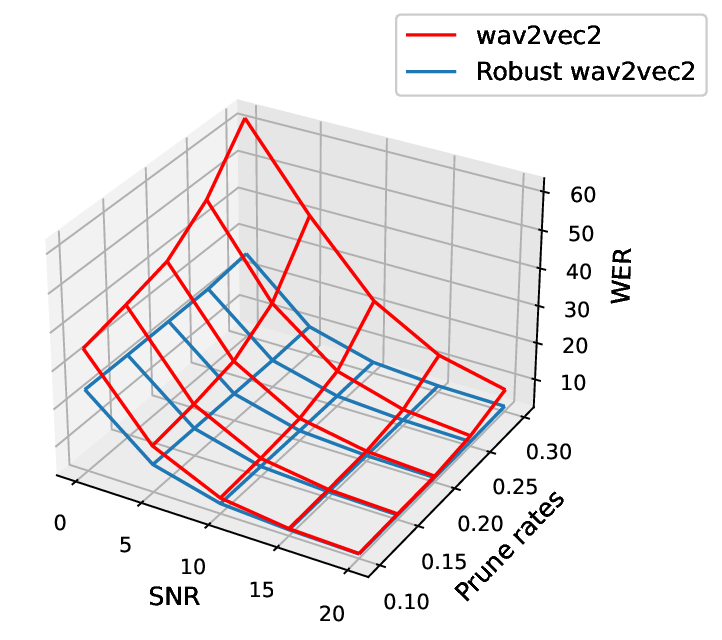}
            \caption{}
            \label{fig:prune_noise_robust}
        \end{subfigure}
        \begin{subfigure}[htpb]{0.49\textwidth}
            \centering
            \includegraphics[width=\textwidth]{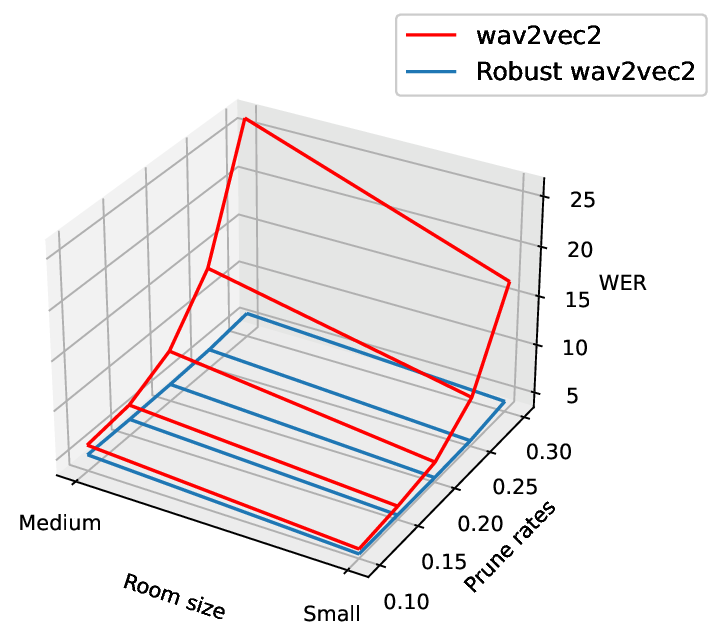}
            \caption{}
            \label{fig:prune_reverb_robust}
        \end{subfigure}
    \caption{WER as a function of the pruning rate and (a)  additive noise or (b) reverberation levels.}
    \label{fig:prune_robust_noise_reverb}
    \end{figure}

Next, we explore the robustness of the models to pruning.  Figures \ref{fig:prune_noise_robust} and \ref{fig:prune_reverb_robust} show WER plots as a function of pruning rate and SNR or pruning rate and room size, respectively. As can be seen, pruning of the two models affected WERs, especially for SNRs lower than 15 dB, with the original wav2vec 2.0 model showing the greatest deterioration, particularly with pruning rates above 20\%.  Reverberation, in turn, showed minimal effect on the pruned robust version, but had a substantial impact on wav2vec 2.0, especially for medium sized rooms and pruning rates greater than 15\%.

\begin{figure}
        \centering
        \includegraphics[scale=0.55]{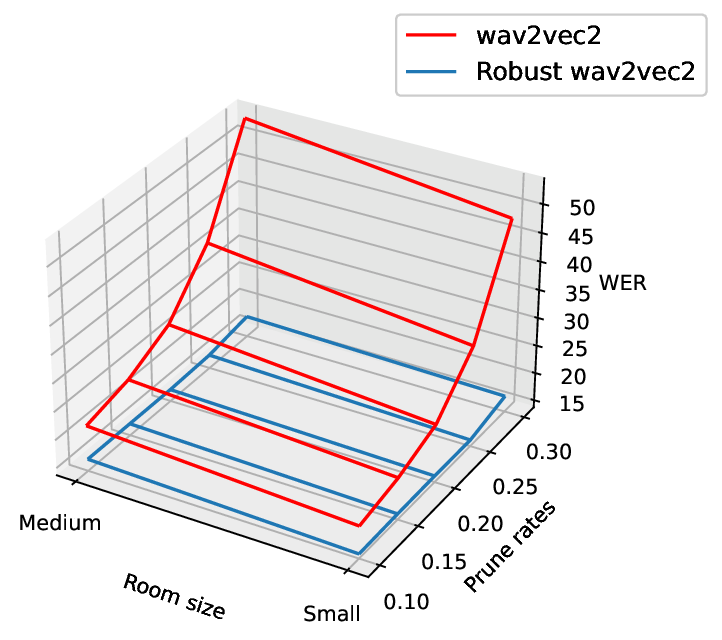}
        \caption{WER as a function of room size for signals with added noise between 0 and 20~dB.}
        \label{fig:prune_noise+reverb_robust} 
\end{figure}

Lastly, we evaluate the robustness of the two models to pruning with combined additive noise and reverberation present in the test signals. Figure \ref{fig:prune_noise+reverb_robust} shows the WER as a function of prune rate and room size where noise levels have been averaged across the 0-20 dB range. Again the robust model showed to be insensitive to increased pruning rates, but sensitive to the degradations themselves. For example, at an SNR of 5 dB, the robust model achieved a WER of 10.1\% at a pruning rate of 0.3. This error increased to 25.5\% at an SNR of 0 dB. Notwithstanding, this is substantially better than what was shown with the original wav2vev 2.0 model that, under the same compression and noise conditions, achieved a WER of 62.2\%.

Overall, the experiments described herein suggest that existing quantization and pruning compression schemes seem to be well-suited for edge speech recognition applications when applied in somewhat clean conditions. In such scenarios, compression rates as high as 4 could be achieved with minimal impact on WER. On the other hand, if test applications involve noisy and/or reverberant conditions, improved speech representations are still needed, beyond what can be achieved with the so-called robust wav2vec model. Environment-aware knowledge distillation may be one possible solution.


\section{Conclusions}
In this work, we evaluate the robustness of two SOTA ASR speech models, namely wav2vec 2.0 and robust wav2vec 2.0, to unseen noisy and reverberant conditions when the models are compressed via quantization and pruning schemes. In particular, 8-bit quantization and L1-norm based global unstructured pruning were explored. It was found that while quantization and pruning have minimal impact on WER in clean conditions, noise and reverberation cause a significant WER degradation, even with models built inherently to be robust to such conditions. Future work should explore more robust compression and self-supervised representations before edge speech recognition applications can be deployed ``in the wild''.

\bibliography{refs}

\begin{thebibliography}{10}
\providecommand{\natexlab}[1]{#1}
\providecommand{\url}[1]{\texttt{#1}}
\expandafter\ifx\csname urlstyle\endcsname\relax
  \providecommand{\doi}[1]{doi: #1}\else
  \providecommand{\doi}{doi: \begingroup \urlstyle{rm}\Url}\fi

\bibitem[Ao et~al.(2021)Ao, Wang, Zhou, Wang, Ren, Wu, Liu, Ko, Li, Zhang, Wei,
  Qian, Li, and Wei]{speechT5}
Junyi Ao, Rui Wang, Long Zhou, Chengyi Wang, Shuo Ren, Yu~Wu, Shujie Liu, Tom
  Ko, Qing Li, Yu~Zhang, Zhihua Wei, Yao Qian, Jinyu Li, and Furu Wei.
\newblock Speecht5: Unified-modal encoder-decoder pre-training for spoken
  language processing.
\newblock 2021.
\newblock \doi{10.48550/ARXIV.2110.07205}.
\newblock URL \url{https://arxiv.org/abs/2110.07205}.

\bibitem[Babu et~al.(2021)Babu, Wang, Tjandra, Lakhotia, Xu, Goyal, Singh, von
  Platen, Saraf, Pino, Baevski, Conneau, and Auli]{speech_XLS-R}
Arun Babu, Changhan Wang, Andros Tjandra, Kushal Lakhotia, Qiantong Xu, Naman
  Goyal, Kritika Singh, Patrick von Platen, Yatharth Saraf, Juan Pino, Alexei
  Baevski, Alexis Conneau, and Michael Auli.
\newblock Xls-r: Self-supervised cross-lingual speech representation learning
  at scale.
\newblock 2021.
\newblock \doi{10.48550/ARXIV.2111.09296}.
\newblock URL \url{https://arxiv.org/abs/2111.09296}.

\bibitem[Baevski et~al.(2020)Baevski, Zhou, Mohamed, and Auli]{wav2vec2.0}
Alexei Baevski, Henry Zhou, Abdelrahman Mohamed, and Michael Auli.
\newblock wav2vec 2.0: A framework for self-supervised learning of speech
  representations.
\newblock \emph{arXiv:2006.11477}, 2020.

\bibitem[Ba et~al.(2016)Ba, Kiros, and Hinton]{layernorm2016}
Jimmy~Lei Ba, Jamie~Ryan Kiros, and Geoffrey~E. Hinton.
\newblock Layer normalization.
\newblock \emph{arXiv:1607.06450}, 2016.
\newblock \doi{10.48550/ARXIV.1607.06450}.
\newblock URL \url{https://arxiv.org/abs/1607.06450}.

\bibitem[Hendrycks and Gimpel(2016)]{gelu2016}
Dan Hendrycks and Kevin Gimpel.
\newblock Gaussian error linear units (gelus).
\newblock \emph{arXiv:1606.08415v4}, 2016.
\newblock \doi{10.48550/ARXIV.1606.08415}.
\newblock URL \url{https://arxiv.org/abs/1606.08415}.

\bibitem[Hsu et~al.(2021)Hsu, Sriram, et~al.]{robusst_wav2vec2.0}
Wei-Ning Hsu, Anuroop Sriram, et~al.
\newblock Robust wav2vec 2.0: Analyzing domain shift in self-supervised
  pre-training.
\newblock \emph{arXiv:2104.01027}, 2021.

\bibitem[Deng et~al.(2020)Deng, Li, Han, Shi, and
  Xie]{model_compression_survey_ieee}
Lei Deng, Guoqi Li, Song Han, Luping Shi, and Yuan Xie.
\newblock Model compression and hardware acceleration for neural networks: A
  comprehensive survey.
\newblock \emph{Proceedings of the IEEE}, 108\penalty0 (4):\penalty0 485--532,
  2020.
\newblock \doi{10.1109/JPROC.2020.2976475}.

\bibitem[Choudhary et~al.(2020)Choudhary, Mishra, et~al.]{Choudhary2020}
Tejalal Choudhary, Vipul Mishra, et~al.
\newblock A comprehensive survey on model compression and acceleration.
\newblock \emph{Artificial Intelligence Review}, 53\penalty0 (7):\penalty0
  5113--5155, Oct 2020.
\newblock ISSN 1573-7462.
\newblock \doi{10.1007/s10462-020-09816-7}.
\newblock URL \url{https://doi.org/10.1007/s10462-020-09816-7}.

\bibitem[Dubey et~al.(2022)Dubey, Gopal, et~al.]{dubey2022icassp}
Harishchandra Dubey, Vishak Gopal, et~al.
\newblock Icassp 2022 deep noise suppression challenge.
\newblock In \emph{Proc. ICASSP}, 2022.

\bibitem[Ko et~al.(2017)Ko, Peddinti, Povey, Seltzer, and
  Khudanpur]{ko2017study}
Tom Ko, Vijayaditya Peddinti, Daniel Povey, Michael~L Seltzer, and Sanjeev
  Khudanpur.
\newblock A study on data augmentation of reverberant speech for robust speech
  recognition.
\newblock In \emph{2017 IEEE International Conference on Acoustics, Speech and
  Signal Processing (ICASSP)}, pages 5220--5224. IEEE, 2017.

\end{thebibliography}

\end{document}